\documentclass[twocolumn, a4paper, 11pt,book]{archive}

\usepackage{version}
\usepackage{graphicx}
\usepackage{rotating}
\usepackage{multirow}
\usepackage{txfonts}

\usepackage{draftwatermark}
\SetWatermarkText{DRAFT}
\SetWatermarkScale{0.75}

\begin{document} 

\twocolumn[{%
 \centering
%
  {\center \bf \huge Non-thermal desorption of complex organic molecules:}\\
\vspace*{0.1cm}
  {\center \bf \LARGE Cosmic-ray sputtering of CH$_3$OH embedded in CO$_2$ ice.}\\
\vspace*{0.2cm}



   {\Large E. Dartois \inst{1}, 
          M. Chabot \inst{2},
         A. Bacmann \inst{3},
         P. Boduch \inst{4},
        A. Domaracka \inst{4},
         H. Rothard \inst{4}
        }\\
\vspace*{0.25cm}
         
   $^1$  Institut des Sciences Mol\'eculaires d'Orsay (ISMO), UMR8214, CNRS, Universit\'e Paris-Saclay, 
B\^at 520, Rue Andr\'e Rivi\`ere, F-91405 Orsay, France\\
              \email{emmanuel.dartois@universite-paris-saclay.fr}\\
  $^2$             Institut de Physique Nucl\'eaire d'Orsay (IPNO), CNRS, Universit\'e Paris-Saclay, F-91406 Orsay, France\\
 $^3$           Univ. Grenoble Alpes, Institut de Plan\'etologie et d'Astrophysique de Grenoble (IPAG), 38000 Grenoble, France\\
  $^4$            Centre de Recherche sur les Ions, les Mat\'eriaux et la Photonique, CIMAP-CIRIL-GANIL, Normandie Universit\'e, ENSICAEN, UNICAEN, CEA, CNRS, F-14000 Caen, France\\
  \vspace*{0.5cm}
{\it \large To appear in Astronomy \& Astrophysics}


  \vspace*{0.5cm} 
 }]

  \section*{Abstract}
   {}
   {Methanol ice is embedded in interstellar ice mantles present in dense molecular clouds. We aim to measure the sputtering efficiencies starting from different ice mantles of varying compositions experimentally, in order to evaluate their potential impact on astrochemical models. The sputtering yields of complex organic molecules is of particular interest, since few mechanisms are efficient enough to induce a significant feedback to the gas phase.}
   {We irradiated ice film mixtures made of methanol and carbon dioxide of varying ratios with swift heavy ions in the electronic sputtering regime. We monitored the evolution of the infrared spectra as well as the species released to the gas phase with a mass spectrometer. 
   Methanol ($\rm^{12}C$) and isotopically labelled $\rm^{13}C$-methanol were used to remove any ambiguity on the measured irradiation products.}
   {The sputtering of methanol embedded in carbon dioxide ice is an efficient process leading to the ejection of intact methanol in the gas phase. We establish that when methanol is embedded in a carbon-dioxide-rich mantle exposed to cosmic rays, a significant fraction (0.2-0.3 in this work) is sputtered as intact molecules. The sputtered fraction follows the time-dependent bulk composition of the ice mantle, the latter evolving with time due to the radiolysis-induced evolution of the bulk. If methanol is embedded in a carbon dioxide ice matrix, as the analyses of the spectral shape of the CO$_2$ bending mode observations in some lines of sight suggest, the overall methanol sputtering yield is higher than if embedded in a water ice mantle. The sputtering is increased by a factor close to the dominant ice matrix sputtering yield, which is about six times higher for pure carbon dioxide ice when compared to water ice.
    These experiments are further constraining
    the cosmic-ray-induced ice mantle sputtering mechanisms important role in the gas-phase release of complex organic molecules from the interstellar solid phase.}
{}


%

\section{Introduction}
Cosmic rays pervade dense clouds and protostellar discs, the resident sites of interstellar ice mantles. They are a source of radiochemistry for these solids, in addition to photolysis from induced secondary vacuum ultraviolet (VUV) 
photons. They are also at the origin of a sputtering process releasing ice grain mantle species and products in the gas phase. This must therefore be considered as a desorption mechanism to be properly quantified to examine to what extent this sputtering in the electronic regime influences the astrochemical balance in dense regions of the interstellar medium.
There is a special interest in the desorption of large molecules embedded in interstellar ice mantles, observed in the gas phase in abundances, often above what classical modelling can predict from pure gas phase chemistry, and even for some including grain chemistry \citep[e.g.][]{Vasyunin2017, Ruaud2016, Bacmann2012, Garrod2008}. At the low temperature of interstellar dust grains, non-thermal chemical, or physical desorption mechanisms can effectively overcome the sublimation barrier.

The first organic molecule considered as 'complex' with significant relative abundances in interstellar ices is methanol. Besides much discussion on its formation pathway, the composition of the ice matrix in which methanol is embedded is not always constrained, and it could vary from source to source \citep[e.g.][]{Bottinelli2010}. The major elements constituting the ice matrix phases are water, carbon dioxide, and carbon monoxide. A carbon monoxide matrix is favoured by several authors on the basis of laboratory experiments showing that the formation of methanol is possible via the hydrogenation of carbon monoxide \citep{Krim2018, Chuang2018, Fuchs2009, Watanabe2002}. Some experiments address alternative formation routes or branching ratios leading to other pathways/species \citep{Qasim2018, Minissale2016, Chuang2016}.
Methanol ice is generally observed in lines of sight with large column densities.
However, no clear correlation emerges between the carbon monoxide content of a cloud and methanol formation, with some lines of sight harbouring low column density upper limits, and others with apparently similar conditions with firm detections (\cite{Whittet2011, Bottinelli2010}). Despite what is commonly assumed in current chemical models, there is no spectroscopic consensus that justifies a strict association of carbon monoxide and methanol formation, with several formation pathways being probably at work.
In several lines of sight, the spectroscopic profiles of the CO$_2$ bending mode that are observed show sub-structures associated with the degeneracy breaking of this mode. It has been experimentally and theoretically demonstrated that such a profile can be assigned to the formation of a methanol/carbon dioxide complex and imply a relatively lower water ice content for this ice phase that would inhibit the complex formation, suggesting a segregation of the ices along these lines of sight \citep[e.g.][]{Klotz2004, Dartois1999, Ehrenfreund1999}. 
The occurrence of high-energy cosmic rays penetrating into these dense regions brings into question their impact on the evolution of these ice mantles and their ability to sputter methanol and more complex organic molecules (COM), returning them to the gas phase.
As ice mantles projected column densities are dominated by water ice, we recently performed experiments focusing on methanol in a water-ice-dominated matrix in \cite{Dartois2019}. 

In this article, we experimentally investigate the sputtering by cosmic rays (swift ions in the electronic regime) of methanol when it is embedded in a carbon dioxide ice matrix. 
In Section 2, the experiments are briefly described. 
Combined infrared and mass spectrometer analyses of the products ejected during swift heavy ion interaction with thin ice films deposited at low temperatures are presented in Section 3. A simple model analysis framework is used to draw figures. The comparison of the methanol sputtering in a carbon dioxide ice matrix with water ice matrix is also discussed. We conclude with the astrophysical implications of methanol sputtering rate.
\section{Experiments}
Swift ion-irradiation experiments were performed at the heavy-ion accelerator Grand Acc\'el\'erateur National d'Ions Lourds (GANIL, Caen, France).
Heavy-ion projectiles were delivered on the IRRSUD beam line\footnote{http://pro.ganil-spiral2.eu/laboratory/experimental-areas}. The Irradiation de GLaces d'Int\'er\^et AStrophysique (IGLIAS) facility, a vacuum chamber (10$^{-9}$ mbar under our experimental conditions) that holds an IR-transmitting substrate that can be cryocooled down to about 10 K, was coupled to the beam line. The ice films were produced by placing the cold window substrate in front of a dosing needle that was connected to the deposition line. Ice films were condensed at 10~K on the window from the vapour phase and were kept at this temperature during the irradiations.
Methanol, purchased from Sigma Aldrich, was of spectrophotometric grade. 
The $\rm^{13}C$ labelled methanol, with a 99\% $\rm^{13}C$ purity was purchased from Eurisotop.
The carbon dioxide was from AirLiquide with  99.995\% purity. They were used as received.
Details of the experimental setup are given in \cite{Auge2018}. 
The stopping power of  the projectiles ($\rm ^{58}Ni^{9+}$ at 0.57 MeV/u) in the electronic regime \citep[calculated with the SRIM package,][]{Ziegler2010} is close to 3.7keV/nm for a pure CO$_2$ ice film, adopting an ice density of 1.1 g/cm$^3$ \citep{Satorre2008}.
For pure methanol, the stopping power is slightly higher, about 5.1keV/nm for a pure ice density  of 1.013g/cm$^3$ \citep[][]{Mate2009}. 
%
%
\begin{table}
\caption{Summary of experiment parameters - $\rm ^{58}Ni^{9+}$ at 0.57 MeV/u}             
\label{table:1}                
\begin{center}
\begin{tabular}{l l l l l}     
\hline\hline       
\#exp.  &T      &N$\rm _0(CO_2)$        &N$\rm _0(CH_3OH)$      &\multirow{2}{*}{$\rm \frac{CH_3OH}{CO_2}$[\%]$\rm^a$}        \\ 
 tag                    &K      &{\tiny[10$^{16}$cm$^{-2}$]}    &{\tiny[10$^{16}$cm$^{-2}$]}                 &                                                                \\ 
 \hline                                      
 {\tiny W2}     &10     &105.4$\pm$1.5                  &7.2$\pm$0.8                    &6.8           \\ 
 {\tiny W3}             &10     &106.2$\pm$1.5                  &32.4$\pm$1.8                &30.5        \\ 
\hline                    
                &       &                                       &N$\rm _0$      &\multirow{2}{*}{$\rm \frac{^{13}CH_3OH}{CO_2}$[\%]}                  \\  
                &       &                                       &$\rm(^{13}CH_3OH)$     &                        \\ 
\hline                    
 {\tiny W1}             &10     &120.6$\pm$1.8                  &6.4$\pm$0.4                    &5.3             \\ 
 {\tiny W1 BIS}         &10     &113.7$\pm$1.2  &13.4$\pm$0.6                 &11.8        \\ 
 {\tiny W1 TER}         &10     &91.0$\pm$1.5            &19.7$\pm$0.7                  &21.6      \\ 
%
%
\hline                  
\end{tabular}
\end{center}
Column density (N$\rm _0$) uncertainties are dominated by the integrated band-strength uncertainties given in Table~\ref{table:band_strengths}; reported uncertainties are the model uncertainty.
For CO$_2$, the antisymmetric stretching mode is used with an adopted integrated absorption cross section of $\rm A=7.6\times 10^{-17}$ cm/molecule. 
For CH$_3$OH, the C-O stretching mode adopted integrated absorption cross section is $\rm A=1.5\times 10^{-17}$ cm/molecule.
$\rm^a$ initial fraction. 
\end{table}
The yield-dependency on the stopping power for pure carbon dioxide ice predicts a total sputtering yield for intact plus radio\-lysed molecules of 5.4$^{+4.3}_{-2.4}$$\times$10$^{4}$ sputtered CO$_2$/ion (using the data shown in \citealp{Rothard2017}).
The ion flux, set between 10$^9$ and 2$\times$10$^9$ ions/cm$^2$/s, was monitored online using the current measured on the beam entrance slits that define the aperture. 
The relation between the current at different slit apertures and the flux was calibrated before the experiments. We used a Faraday cup that was inserted in front of the sample chamber to do this. 
The deposited ice-film thicknesses allowed the ion beam to pass through the film with an almost constant energy loss per unit path length.
A Bruker Fourier Transform InfraRed (FTIR) spectrometer (Vertex 70v) with a spectral resolution of 1 cm$^{-1}$ was used to monitor the infrared (IR) film transmittance. The evolution of the IR spectra was recorded as a function of the ion fluence.
The irradiation was performed at normal incidence, whereas the IR transmittance spectra were recorded simultaneously at 12$\rm^o$ of incidence (a correction factor of $\approx$0.978 was therefore applied to determine the normal column densities).
A sweeping device allows for uniform and homogeneous ion irradiation over the target surface.
Mass spectrometry measurements were performed simultaneously using a microvision2 mks quadrupole mass-spectrometer (QMS). The QMS signals were noise-filtered to smooth out high-frequency temporal fluctuations using a Lee filter algorithm with a typical box size of five to seven points.
The mass-to-charge (m/z) ranges that we scanned were varied for the different experiments in order to optimise the integration time.
The m/z channels we used to follow a given species were selected in order to avoid overlap with channels that are dominated either by contamination or to avoid confusion with a byproduct of the radiolysis.
Labelled $\rm^{13}C-$methanol was chosen so that the clean m/z=33 can be used to follow unambiguously $\rm^{13}CH_3OH^+$ with the QMS. Methanol ($\rm^{12}C$) was also used in the first experiments, in order to follow also the ice mixture infrared spectral behaviour expected in space. Carbon dioxide can be monitored
using m/z=44 (CO$_2^+$).
When only selected optimum m/z are used, a correction factor needs to be applied to estimate the abundance of a species so that the mass-fragmentation pattern following electron impact ionisation can be taken into account. The experimental fragmentation pattern within our QMS for a species X was monitored during the injection of the gas mixture, that is, when the ice film was deposited. A self-calibration of the QMS for the mass-fragmentation pattern of species X was then obtained at m/z as follows:
\begin{equation}
\rm f(X,m/z)=I(m/z)/\sum_{m/z=X_{fragments}}I(m/z)
,\end{equation}
where only the expected m/z from possible (major) fragments (X$\rm _{fragments}$) were included. For $\rm^{13}$CH$_3$OH, the main m/z are 30($\rm H^{13}CO^+$), 31($\rm H_2^{13}CO^+$), 32($\rm^{13}CH_3O^+$), and 33($\rm^{13}CH_3OH^+$).
For CO$_2,$ the main m/z are 12($\rm C^+$), 16($\rm O^+$), 22($\rm CO_2^{2+}$), 28($\rm CO^+$), and 44($\rm CO_2^+$).
Therefore, the sum of the chosen m/z was used and divided by their relative fragmentation-pattern percentage (obtained for this QMS, as explained above) to retrieve the total number of species X.
The spectra were also corrected for the total electron-impact ionisation cross section $\rm \sigma^{impact}(X)$ at 70 eV (energy of the QMS electron ionisation source) for each molecule: CO$_2$ (3.521~$\AA^2$, NIST database), CH$_3$OH \citep[4.8~$\AA^2$,][]{Vinodkumar2011}, also used for $\rm^{13}$CH$_3$OH, and for one main radiolysis product used in the analysis, CO (2.516~$\AA^2$, NIST database), to compensate for the higher ionisation efficiency of larger species (which carry more electrons).

The abundance ratios of species X and Y were thus evaluated from
\begin{equation}
\rm \frac{[X]}{[Y]}=
\frac{\sum_{m/z=X_{fragments}}I(m/z)}{\sum_{m/z=X_{fragments}}f(X,m/z)} 
\frac{\sum_{m/z=Y_{fragments}}f(Y,m/z)}{\sum_{m/z=Y_{fragments}}I(m/z)}
\frac{\sigma^{impact}(Y)}{\sigma^{impact}(X),}
\end{equation}
with the fragments chosen so that they have a significant signal-to-noise ratio and do not overlap with other {dominant} species and/or potential residual gas m/z
contribution. 
%
%
\begin{figure*}[tbhp]
\begin{minipage}[c]{\columnwidth}
\includegraphics[width=\linewidth]{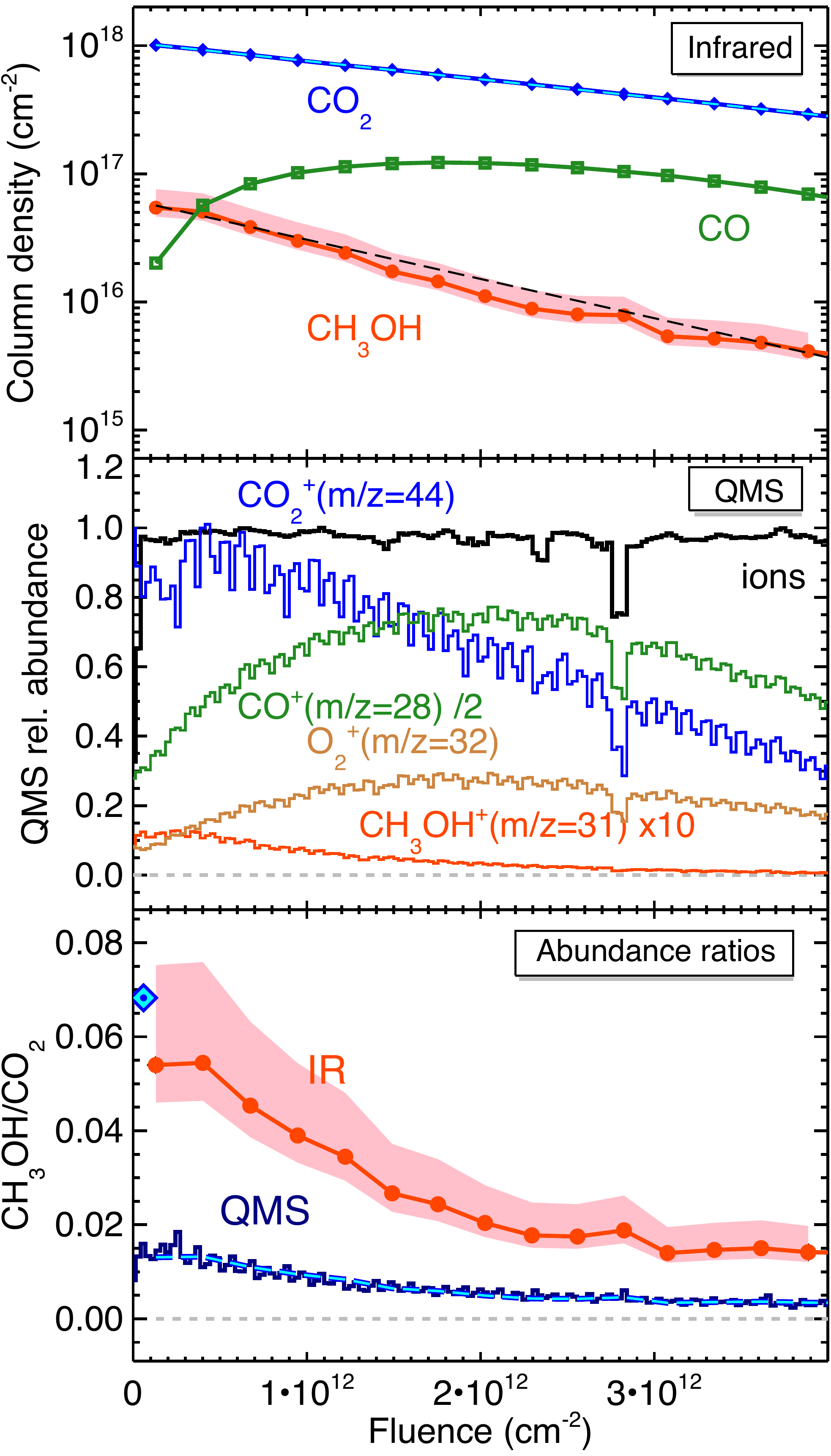}
\end{minipage}
\begin{minipage}[c]{\columnwidth}
\includegraphics[width=\linewidth]{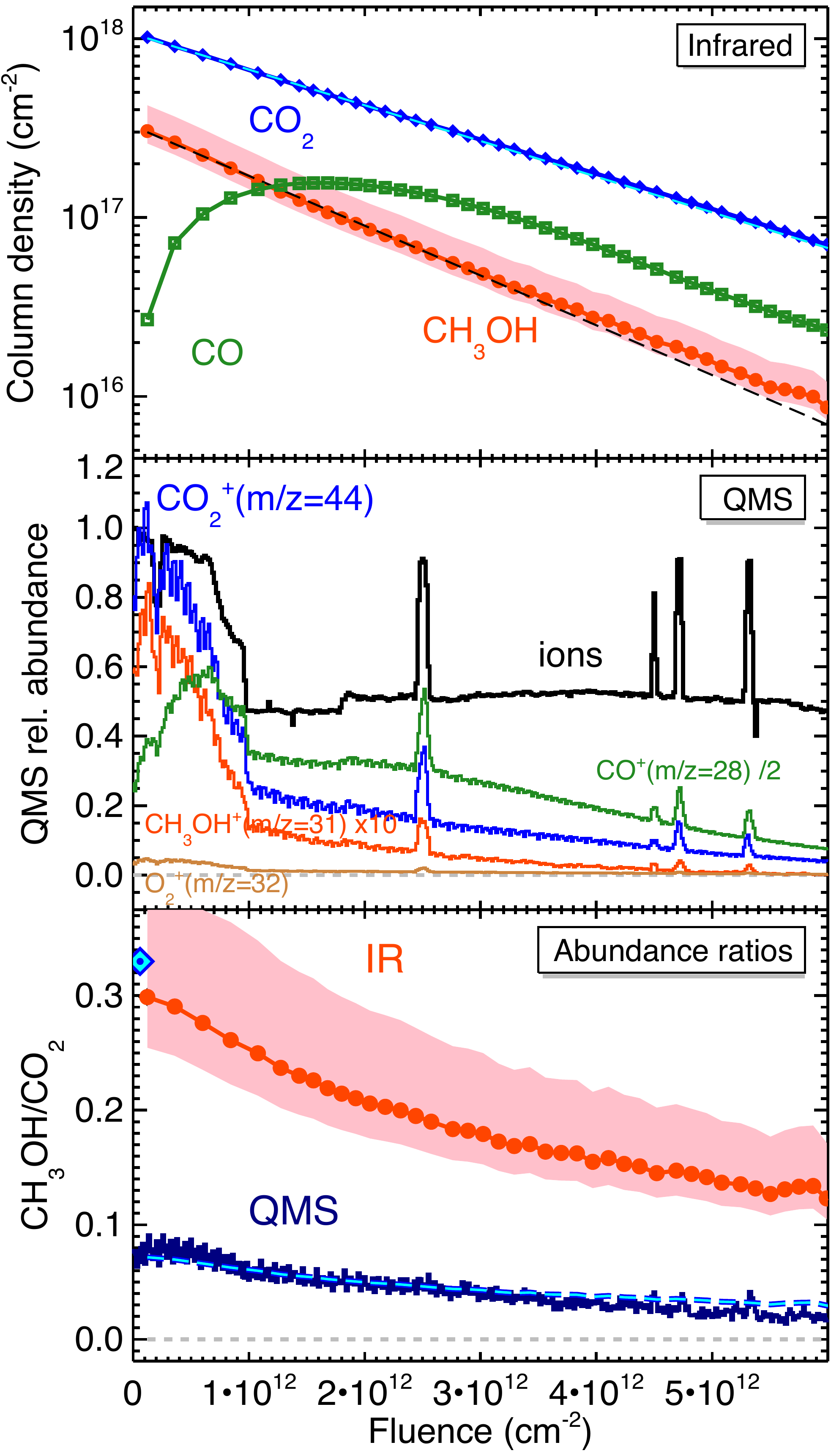}
\end{minipage}
\caption{Methanol carbon dioxide ice-film mixture experiments. Left: CH$_3$OH/CO$_2$ (6.8\%). Right: CH$_3$OH/CO$_2$ (30.5\%).
Upper panels: Ice column-density measurements from IR spectra. The column densities are estimated using the integrated cross sections from Table~\ref{table:band_strengths}. The cyan and black dashed lines represent the fits to the destruction cross section of carbon dioxide and methanol, respectively.
Middle panels: QMS-normalised signals used to follow the relative abundance of carbon dioxide (m/z=44), methanol (m/z=31; m/z=32 cannot be used, as it is strongly overlapping with dioxygen), and carbon monoxide (m/z=28, after subtracting the contribution of the mass fragmentation of CO$_2$) during irradiation. The black line is the scaled monitoring signal of the ion flux, showing its stability. 
The m/z=28 signal increases at the beginning, because it contains fragments from the radiolytic products of the ice mixture.
Lower panels: CH$_3$OH/CO$_2$ abundance ratio deduced from the IR spectra of the ice film as a function of fluence (red dots, uncertainty filled in in red). Comparison with the QMS-determined abundance ratio of the same desorbed molecules (dark blue line, from m/z=31 for CH$_3$OH and m/z=44 for CO$_2$). The blue and cyan diamond symbol indicates the gas ratio measurement with the QMS during the ice film deposition. The dashed blue and cyan line represents $\rm\chi= \beta_{meas} {\scriptstyle\times} ({f_{CH_3OH}}/{f_{CO_2}})^{bulk}$. See text for details.}
\label{fig:mosaic_12C}
\end{figure*}
%
%
%
%
%
\begin{sidewaysfigure*}[tbhp]
    \centering
\begin{minipage}[c]{0.33\columnwidth}
\includegraphics[width=\linewidth]{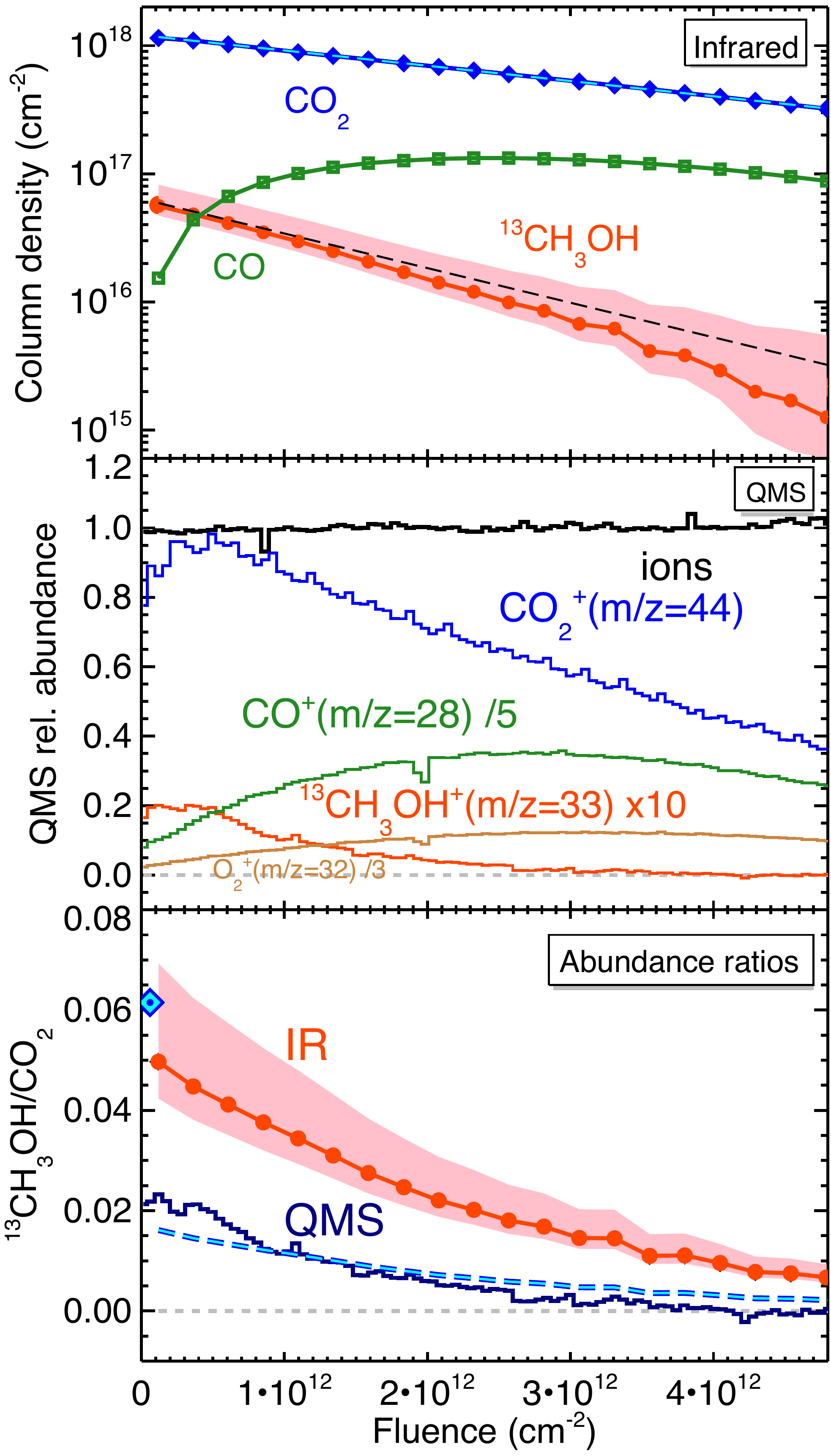}
\end{minipage}
\begin{minipage}[c]{0.33\columnwidth}
\includegraphics[width=\linewidth]{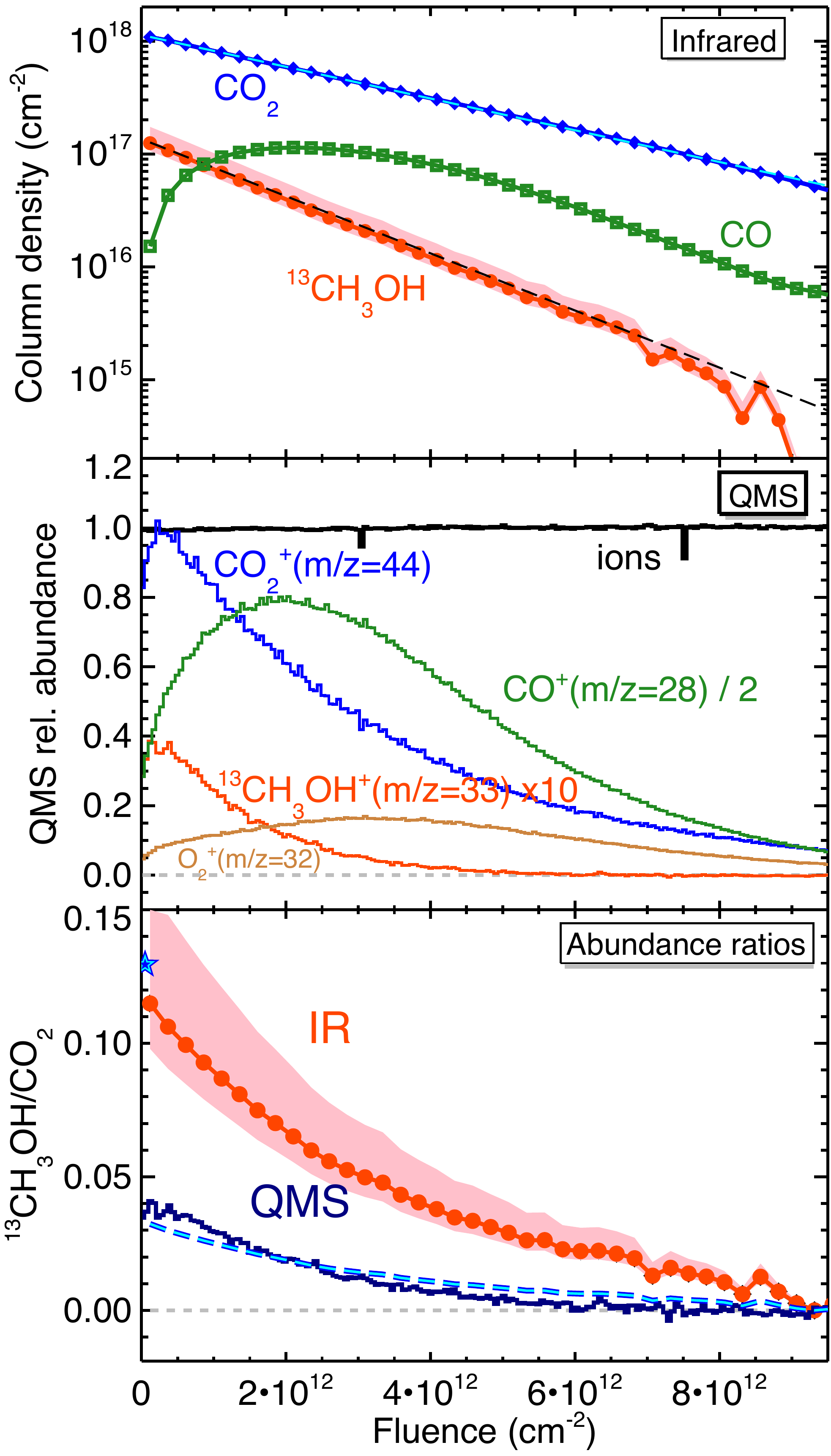}
\end{minipage}
\begin{minipage}[c]{0.33\columnwidth}
\includegraphics[width=\linewidth]{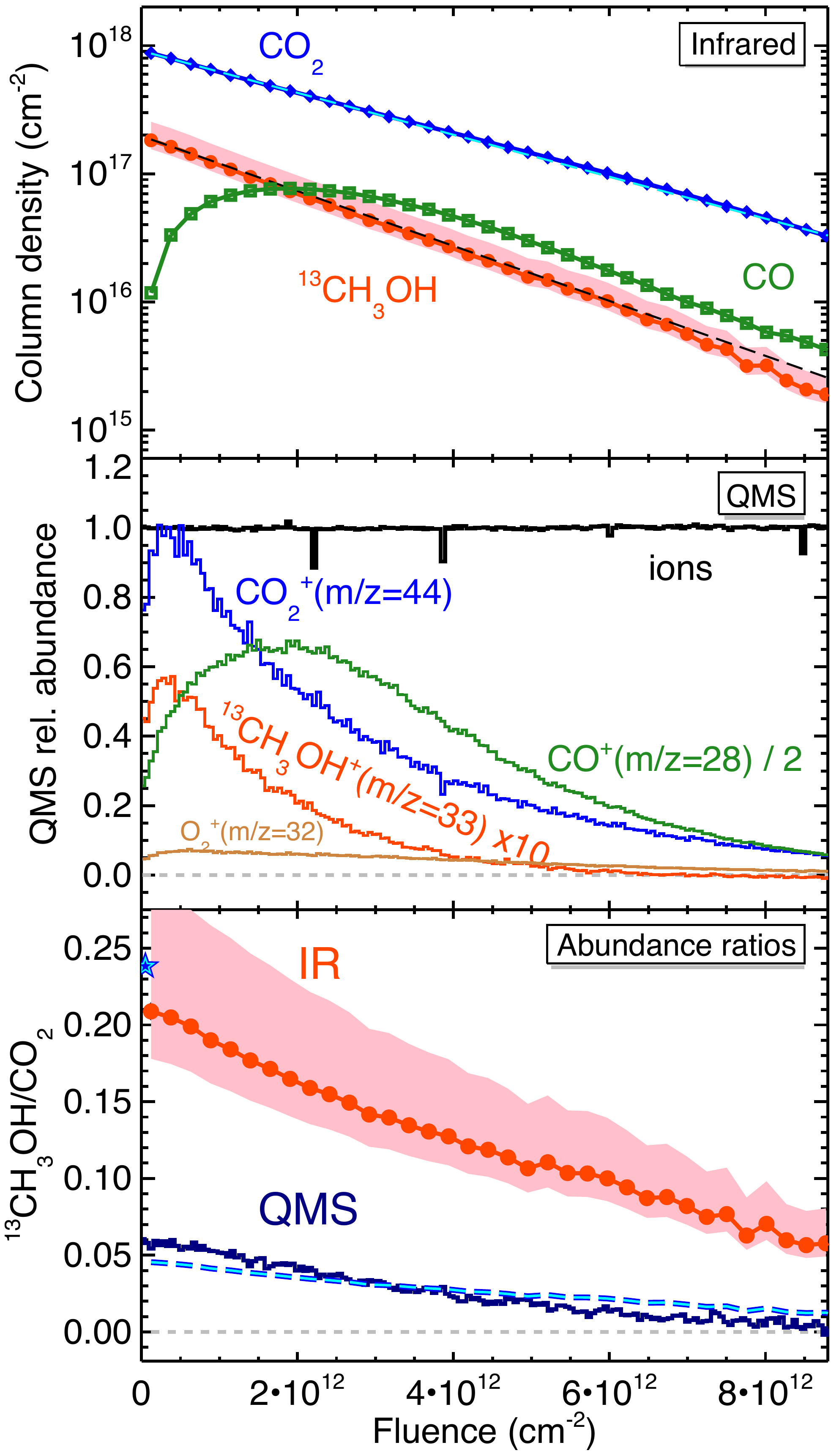}
\end{minipage}
\caption{Same as Fig.\ref{fig:mosaic_12C} for $\rm^{13}C-$methanol carbon dioxide ice-film mixtures. 
From left to right: CH$_3$OH/CO$_2$ (5.3\%, 11.8\%, 21.6\%).
$\rm^{13}C$-methanol is followed using the radiolytic and fragmentation pattern free m/z=33 QMS channel. It should be noted that the maximum fluence for the left panel experiment is only about half that  of the others, as the initial methanol fraction is low and no longer detectable in the IR for fluences above about $\rm 4\times10^{12}$~cm$^{-2}$.
See text for details.}
\label{fig:mosaic_13C}
\end{sidewaysfigure*}
%
%
In the mixture experiments, in order to establish the relative abundance of species, the QMS signals are presented corrected for the relative mass-fragmentation pattern for a given mass attributed to a dominant species, normalised by its electron-impact ionisation cross section, as discussed above. The signals are then scaled by a global factor.
During the experiments, when the ion beam was stopped, we recorded and followed the evolution of the QMS and chamber background signal.
The QMS spectra we present are background subtracted.
A summary of the ice-film parameters, such as ice mixture and irradiation temperatures, is given in Table~\ref{table:1}.
A summary of the integrated band strengths from the literature used in the IR spectra analysis to determine ice-film column densities is listed in Table~\ref{table:band_strengths}.
\section{Results}
\label{results}
The outcome of the irradiation experiments performed on methanol or $\rm^{13}C$-methanol isotopically labelled ices mixed in various proportions with carbon dioxide are reported below. The usefulness of the $\rm^{13}C$-methanol isotopologue in the yield determination is emphasised. The sputtering yield is analysed with a model including the evolution of the ice composition upon ion irradiation. 
\subsection{CH$_3$OH:CO$_2$ ice mixtures}
Mixtures of methanol and carbon dioxide ice with two different proportions CH$_3$OH/CO$_2$ ($\approx$0.07 and 0.3) were irradiated. The recorded IR and QMS evolution in function of the ion fluence are shown in Fig.~\ref{fig:mosaic_12C}. 
Using the C-O stretching mode of methanol and the CO$_2$ antisymmetric stretching mode, we report the column density of these molecules in the ice.
The abundance ratio of gaseous methanol to carbon dioxide measured with the QMS during the ice deposition (blue diamonds in Fig.\ref{fig:mosaic_12C}, lower panels) gives, within uncertainties, the same value as the ratio measured after deposition for the ice film in the IR (and providing therefore an independent measurement of the expected deposited ice-mixture ratio).
The abundance ratios of gas- and solid-phase methanol to carbon dioxide evolve in parallel during irradiation. The measured abundance ratio in the gas phase is lower than measured in the bulk of the ice film, showing that the methanol sputtering yield is influenced both by sputtering and by higher radiolytic destruction efficiencies.
In the figures, we also provide scaled ion flux variations to show the ion beam variability. Note that the ion beam was fluctuating significantly during the second experiment, and thus the magnitude of the species in the QMS signals. However, the abundance ratios calculated are largely unaffected, only marginal correlation exists with the fluctuations.
The mass spectra in Fig.\ref{fig:mosaic_12C} show that carbon dioxide and methanol are desorbed as soon as the ion irradiation begins (middle panels) and decreases (if corrected with beam fluctuations, this is also the case for the second experiment).
We note that, at the very beginning of the irradiation, for fluences lower than a few 10$^{11}$ cm$^{-2}$, a slight phase change can occur with the first ions impinging the freshly deposited ice film, involving a restructuring of the ice. During this early phase, a steady state is not reached in the chamber, the ice volume can change by up to about ten percent
; therefore, the sputtering efficiency can be slightly altered.
For the m/z=28 (gas-phase CO) and m/z=32 (dominated by O$_2$), the intensities are lower at the beginning of the irradiation, then rise, and finally, decrease. This behaviour is a way to distinguish the direct products from the accumulated bulk ones. 
The observed gaseous CO arises partly due to direct sputtered CO$_2$ and methanol radiolytic product (value at the very beginning of the irradiations), and partly from the bulk of the ice at later times (bell-shaped curve behaviour).
In addition, as it is more volatile, the desorbing CO can come from deeper layers of the ice film than carbon dioxide and methanol. This is shown by the amount of m/z=28 that desorbs with respect to methanol, whereas the mean bulk ratio of these ices is lower (upper panel of Fig.\ref{fig:mosaic_13C}) than what the QMS observes.
The extent of the radiolysis of carbon dioxide and methanol with respect to the sputtering of intact molecules 
is discussed in a following section.

\subsection{$^{13}$CH$_3$OH:CO$_2$ ice mixtures}
Mixtures of $\rm^{13}C$-labelled methanol and carbon dioxide ice with varying proportions ($^{13}$CH$_3$OH/CO$_2$$\approx$0.05, 0.12, 0.22) were irradiated. The recorded IR and QMS evolution in function of the ion fluence are shown in Fig.~\ref{fig:mosaic_13C}. The benefit of using an isotopic marker is that it provides a clean channel for the QMS measurements at m/z=33 to directly follow $\rm^{13}CH_3OH^+$ with neither disturbance from radiolysis products of methanol nor CO$_2$. Otherwise, the description is the same as for the non-labelled methanol experiments discussed above, and we confirm the behaviour observed. 
%
%
%
\begin{table*}
\caption{Results}             
\label{table:2}                
\begin{center}
\begin{tabular}{l l l l l l }     
\hline\hline       
\#exp.          &$\rm \sigma_{CO_2}$            &$\rm \sigma_{CH_3OH}$  &$\rm N_d$$\rm^a$                     &$\rm \beta_{calc}$$\rm^b$              &$\rm \beta_{meas}$$\rm^c$\\
 tag            &{\tiny[10$^{-13}$cm$^{2}$]}    &{\tiny[10$^{-13}$cm$^{2}$]}    &{\tiny[10$^{16}$cm$^{2}$]}     &                                                       &\\
\hline                                      
 {\tiny W2}     &3.3$\pm$0.1                            &7.1$\pm$3.0                            &5.88$^{+4.68}_{-2.61}$         &0.36$^{+0.35}$                         &0.24$\pm$0.03\\
 {\tiny W3}     &4.6$\pm$0.1                            &6.4$\pm$0.4                            &4.67$^{+3.72}_{-2.08}$         &0.74$^{+0.14}_{-0.73}$                 &0.24$\pm$0.02\\
\hline                    
                        &                                               &$\rm \sigma_{^{13}CH_3OH}$   &                                               &                                                       &\\
\hline                    
 {\tiny W1}             &2.7$\pm$0.1    &6.2$\pm$0.5    &7.08$^{+5.64}_{-3.15}$ &0.29$^{+0.39}$         &0.32$\pm$0.09\\
 {\tiny W1 BIS}         &3.2$\pm$0.1    &5.8$\pm$0.3    &6.77$^{+5.39}_{-3.01}$ &0.45$^{+0.31}$         &0.28$\pm$0.06\\
 {\tiny W1 TER}         &3.8$\pm$0.1    &4.9$\pm$0.2    &5.63$^{+4.48}_{-2.50}$ &0.79$^{+0.11}_{-0.55}$ &0.22$\pm$0.05\\
%
%
\hline                  
\end{tabular}
\end{center}
Destruction cross sections ($\rm \sigma_{CO_2}$ and $\rm \sigma_{CH_3OH}$) uncertainties are dominated by the integrated band-strength uncertainties given in Table~\ref{table:band_strengths}; reported uncertainties are the model uncertainty.
 $\rm^a$ The depth of desorption ($\rm N_d$) comes from equation~\ref{equation_Nd}. $\rm^b$ $\rm \beta_{calc}$ is obtained from equation~\ref{equation_QMS}. A single upper error bar means there is only an upper bound. $\rm^c$ The $\rm \chi = \beta_{meas} {\scriptstyle\times} ({f_{CH_3OH}}/{f_{CO_2}})^{bulk}$ value is shown as dashed blue and cyan lines in lower panels of Fig.~\ref{fig:mosaic_12C} and \ref{fig:mosaic_13C}.
\end{table*}
%
%
\subsection{Sputtering yield evaluation \label{section_sputtering}}
%
A model of the evolution of column densities for different species with the projectile ion fluence as monitored by infrared spectroscopy for the ice (bulk), and a first-order model to estimate the abundance ratio measured in the gas phase by the QMS, were explained in detail in \cite{Dartois2019}. Here, we recall this simple model describing the fraction of intact sputtered molecules $\rm \chi$ in the gas phase. For instance, the abundance ratio measurements with the QMS can be expressed as
\begin{align}
\rm \chi = \left(\frac{CH_3OH}{CO_2}\right)^{QMS}       & \rm  \approx \frac{Y_{eff}-\sigma_{CH_3OH}~N^d}{Y_{eff}-\sigma_{CO_2}~N^d} \left(\frac{f_{CH_3OH}}{f_{CO_2}}\right)^{bulk,}  \nonumber\\
& \rm = \beta \left(\frac{f_{CH_3OH}}{f_{CO_2}}\right)^{bulk} \label{equation_QMS}
\end{align}
with $\rm f_{CH_3OH}$ and $\rm f_{CO_2}$ being the fraction of methanol and carbon dioxide molecules in the sputtered ice layers, respectively (i.e. their fraction in the ice composition), whereas $\rm \sigma_{CH_3OH}$ and $\rm \sigma_{CO_2}$ are their destruction cross sections.
Adopting a cylindrical geometry for the sputtered volume \citep[e.g.][Fig.1]{Dartois2018}, $\rm N^d$ corresponds to the column density of molecules in the ice film at the depth of desorption (i.e. the height of the cylinder). $\rm Y_{eff}$ is the semi-infinite thickness effective-sputtering 
total yield for the ice mixture under study. A total yield means that it includes the intact and radiolysed molecules sputtered from the volume considered. Unlike water ice, which is relatively resilient against destruction \citep[e.g.][and references therein]{Dartois2019}, the carbon dioxide destruction cross section becomes important in the evaluation of the yield, and it cannot be neglected. To this simple model, we add an additional radiolysis efficiency factor, called $\rm \eta_{CO_2}$, for carbon dioxide.
In an ice matrix dominated by CO$_2$, the absolute intact carbon dioxide sputtering yield is therefore
\begin{align}
\rm Y^{CO_2}_{abs} =  Y_{eff}^{CO_2} \, \eta_{CO_2,}
\label{equation_net}
\end{align}
$\rm \eta_{CO_2}$ being the intact fraction of sputtered carbon dioxide molecules when compared to the sum of the intact and products species sputtered. The by far dominant radiolytic product of CO$_2$ is CO (see Fig~\ref{fig:fragmentation_patterns_CH3OH_CO2}-\ref{fig:fragmentation_patterns_13CH3OH_CO2}), therefore
\begin{align}
\rm \eta_{CO_2} \approx \left(\frac{CO_2}{CO_2+CO}\right)^{QMS.}
\label{equation_net}
\end{align}
The CO value is evaluated at early stages of the ion irradiation, when CO has not yet accumulated in the ice. The $\eta$ value is, threrefore, probably underestimated, and more intact CO$_2$ molecules are sublimating, since CO is more volatile, and some CO contribution can come from deeper layers than CO$_2$.
This intact fraction can be related to the processes occurring in the solid phase by
\begin{align}
\rm \left(\frac{CO_2}{CO_2+CO}\right)^{QMS} \approx \left(\frac{Y_{eff}^{CO_2}-\sigma_{CO_2}~N^d}{Y_{eff}^{CO_2}}\right)^{bulk,} 
\label{equation_net}
\end{align}
where the superscript 'bulk' indicates that these values are measured with the evolution of the ice film in the infrared.
We experimentally measured
$\rm \sigma_{CO_2}$. $\rm Y_{eff}^{CO_2}$ can be evaluated from infrared measurements.
Then we can infer 
\begin{align}
\rm N^d \approx \frac{Y_{eff}^{CO_2}(1- \eta_{CO_2})}{\sigma_{CO_2.}} 
\label{equation_Nd}
\end{align}
The radiolytic destruction cross sections $\rm \sigma^{destruction}_{CO_2}$ of carbon dioxide, and $\rm \sigma^{destruction}_{CH_3OH}$ of methanol, are obtained in each experiment by fitting the column density evolution as in function of fluence using IR measurements (the films used in this study are thick enough for the sputtering not to contribute significantly to the IR evolution).
The fitted cross-section curves are shown overlaid as dashed lines in the upper-right panels of Fig.\ref{fig:mosaic_12C} and \ref{fig:mosaic_13C}.
In the case of a carbon dioxide dominated ice mantle, the effective sputtering yield should be close to the yield of carbon dioxide ice. 
The sputtering yield with heavy ions in the MeV/u range is calculated from previous measurements \citep[Ni,Xe,Ti;][]{Seperuelo2009,Mejia2015,Rothard2017}. Assuming a quadratic dependency with the electronic stopping power, we calculate that $\rm Y_{eff}^{CO_2}=5.4^{+4.3}_{-2.4}{\scriptstyle\times}10^{4}$ CO$_2$/ion. 

An approximate value for $\rm N_d$ can therefore be estimated from equation~\ref{equation_Nd}, and used in equation~\ref{equation_QMS} to infer the value of the proportionality factor $\rm\beta_{calc}$. This factor is reported in table~\ref{table:2}. The calculated values of $\rm\beta_{calc}$ are in fair agreement, taking into account the large uncertainties, with the value measured in our experiments. The $\rm \chi=\beta_{meas} {\scriptstyle\times} ({f_{CH_3OH}}/{f_{CO_2}})^{bulk}$ values are shown as dashed blue and cyan lines in the lower panels of Fig.~\ref{fig:mosaic_12C} and \ref{fig:mosaic_13C}. 
We used an isotopic marker for methanol in some ice mixtures, providing a clean channel (m/z=33) for its QMS detection and measurements, ensuring a safer $\rm \beta_{meas}$ determination. The ice mixtures explored, with $\rm^{12}C$ or $\rm^{13}C$ isotopically enriched methanol, give consistent results for the sputtering yields, as shown in table~\ref{table:2}.
\section{Discussion}
\label{discussion}

\subsection{Absolute sputtering rate}
According to the beta value we determined, about one third of the methanol-to-carbon-dioxide ratio observed in our experiments in the ice (bulk) is ejected by sputtering. To obtain the absolute methanol sputtering yield, this value must be multiplied by $\rm \eta_{CO_2}$, the ejected intact CO$_2$ fraction, which is of the order of sixty percent. As a consequence, overall, when methanol is within a CO$_2$ rich ice mixture, about twenty percent of the methanol present in the ice is desorbed intact. 
Assuming that the fraction of dissociated molecules does not depend strongly on the stopping power, we calculated the astrophysical sputtering rate in the same way as discussed in \cite{Dartois2015}, with an effective total yield proportional to the square of the electronic stopping power. We obtain a rate of 83.8 $\rm cm^{-2}s^{-1}$ for a pure CO$_2$ ice for a reference cosmic-ray-ionisation rate of $\rm \zeta=10^{-16}s^{-1}$, therefore, the absolute methanol rate is about $\rm \approx16.8{\scriptstyle\times}({f_{CH_3OH}}/{f_{CO_2}})^{bulk}$ $\rm cm^{-2}s^{-1}(\zeta/10^{-16}s^{-1}$). 
For the pure water ice case, the equivalent sputtering rate would be 13.9 $\rm cm^{-2}s^{-1}$ for a reference $\rm \zeta=10^{-16}s^{-1}$. According to equations 6 and 7 in \cite{Dartois2019}, we expect between about 30 to 40\% of the methanol-to-water fraction to be desorbed by the sputtering process in the electronic regime. The fraction radiolysed in a methanol-CO$_2$ mixture is slightly higher, 
as can be seen from the values of the $\rm \beta_{meas}$ in Table~\ref{table:2}, but the higher sputtering efficiency of the CO$_2$ matrix by a factor of about six largely dominates the net sputtering rate.

The relation between the expected secondary VUV photon flux and the cosmic-ray ionisation rate, under dense-cloud conditions where ices predominate and when the external UV field is fully attenuated, is discussed in, e.g., \cite{Shen2004,Prasad1983}. Based on these works, we adopt a value of 
7940 VUV $\rm photons.cm^{-2}s^{-1}$ for $\rm \zeta=10^{-16}s^{-1}$.
To match the cosmic-ray electronic sputtering efficiency, the absolute photodesorption yield should be of the order of $\rm \approx2.1{\scriptstyle\times}10^{-3}({f_{CH_3OH}}/{f_{CO_2}})^{bulk}$/VUV photon for methanol/carbon dioxide mixture in the range explored in this study. 
Photodesorption measurements for methanol in the literature include pure methanol-ice desorption. \cite{Bertin2016}, and \cite{Cruz-Diaz2016}, deduced rates of the order of  $10^{-5}$/VUV photon for the former, and an upper limit of about $\rm <3\times10^{-5}$/VUV photon for the latter.
\cite{Bertin2016} also studied a methanol/CO mixture, and the rate they inferred falls to less than $10^{-6}$ molecules/photon, meaning lower than pure methanol, whereas pure CO photodesorbs more efficiently. One cannot infer directly from these measurements what would be the case for a methanol/CO$_2$ mixture (as explored in this article) but the lowering of the effective methanol-photodesorption rate for the methanol/CO mixture probably betrays the fact that mixtures open new photodissociation/recombination routes. Methanol/CO$_2$ mixtures may thus also give a lower photodesorption rate value than pure methanol.

\subsection{Methanol position-FWHM diagram}

\begin{figure}
\centering
\includegraphics[width=\linewidth]{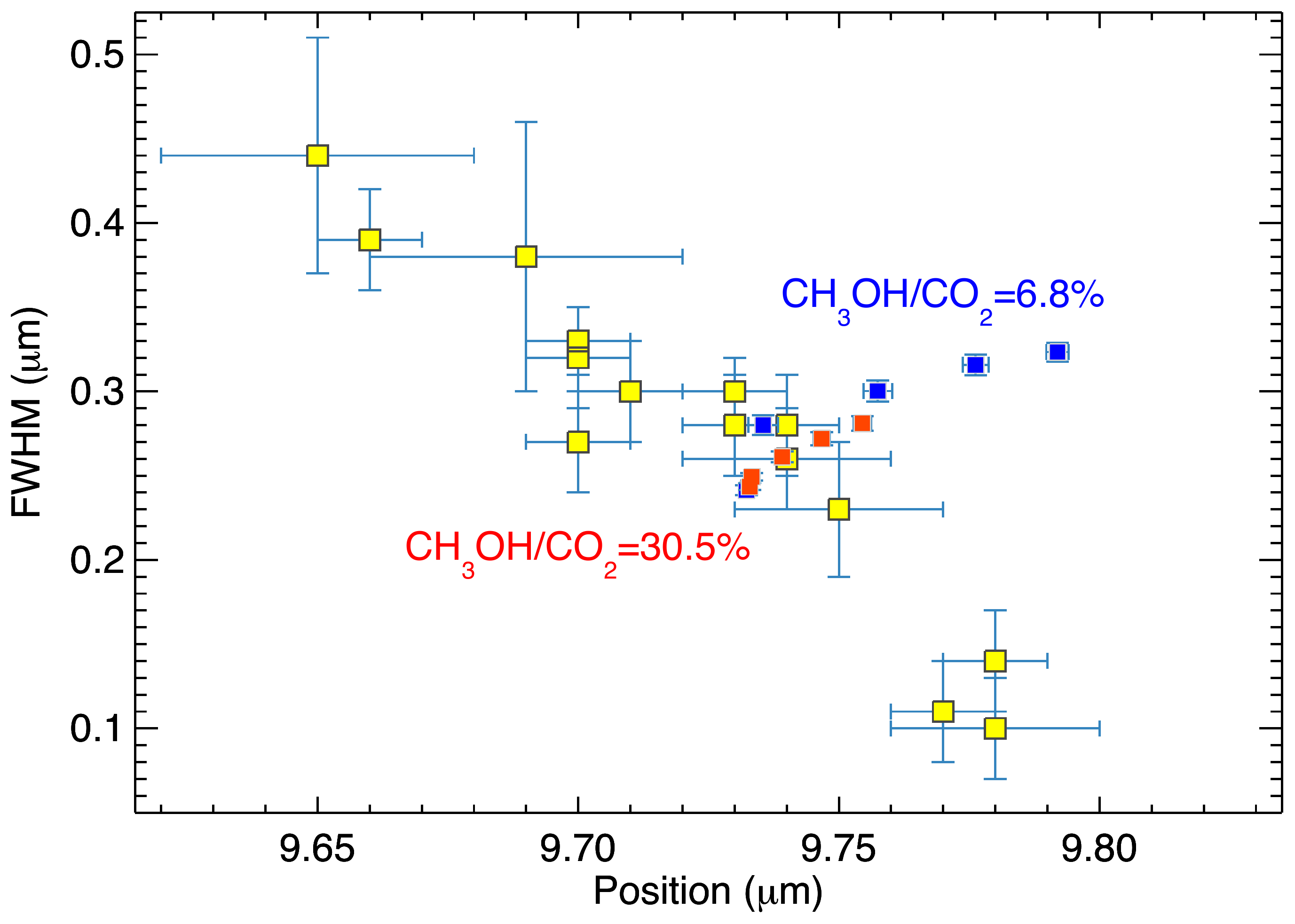}
\caption{Methanol C-O stretching-mode position FWHM diagram of the experiments performed in this work, compared to ISM measurements presented in \cite{Bottinelli2010} (yellow squares). See text for details.
}
\label{fig:positions}
\end{figure}

Methanol was investigated with Spitzer space telescope observations in dense regions \citep[][]{Bottinelli2010}.
Botinelli and colleagues showed in their Fig.12 that most of the observed methanol C-O stretching-mode position versus full width at half-maximum (FWHM) is in a mantle whose composition is dominated neither by a H$_2$O-rich nor by a CO-rich matrix. The methanol C-O stretching-mode position versus FWHM diagram for our experiments with $\rm^{12}C-$methanol/carbon dioxide mixtures are reported in Fig.\ref{fig:positions}, for the first infrared measurements during irradiation.  As the irradiation proceeds, the positions shift progressively towards higher FWHM and redder centroid positions. At this stage, the ice mixtures are heavily irradiated and no longer defined by the initial methanol/carbon dioxide mixtures. The ice mixtures used initially, and their evolution upon moderate irradiation fluences, are also compatible spectroscopically with the observed astronomical positions and 
FWHM values.
%
%
%
\section{Conclusions}

We measured the sputtering yield of methanol embedded in a carbon dioxide ice matrix at 10K irradiated by swift heavy ions in the electronic regime of energy deposition, simulating experimentally interstellar cosmic rays.
We conclude that a large fraction of intact molecules are desorbed by cosmic rays with a sputtering yield close to that of the carbon dioxide ice matrix. A significant fraction of the sputtered CO$_2$ is also radiolytically processed, leading to the ejection of fragments (dominated by CO). The fraction of carbon dioxide molecule sputtered as products must be taken into account to obtain the absolute sputtering yield.
The methanol-to-carbon-dioxide ratio observed in the gas phase by mass spectrometry is proportional to the bulk ice mantle composition, as measured with infrared spectroscopy. The proportionality factor is about one third in the experiments presented here, which reflects the fact that the destruction cross section of methanol is higher than that of the carbon dioxide one when exposed to cosmic rays. The overall efficiency of the initial methanol present  in the ice mantle sputtered intact is about twenty percent. 
The higher efficiency of sputtering of the carbon dioxide ice matrix as compared to more bounded systems, such as a water-ice-dominated ice mantle, imply a very efficient release of complex organic molecules such as methanol.

\begin{acknowledgements}
This work was supported by the Programme National "Physique et Chimie du Milieu Interstellaire" (PCMI) of CNRS/INSU with INC/INP co-funded by CEA and CNES, by the P2IO LabEx program: "Evolution de la mati\`ere du milieu interstellaire aux exoplan\`etes avec le JWST" 
and the ANR IGLIAS, grant ANR-13-BS05-0004 of the French Agence Nationale de la Recherche.
Experiments performed at GANIL. We thank T. Madi, T. Been, J.-M. Ramillon, F. Ropars and P. Voivenel for their invaluable technical assistance. 
We would like to acknowledge the anonymous referee for constructive comments that significantly improved the content of our article, as well as the editor M. Tafalla and language editor Natasha SG. 
\end{acknowledgements}


%
%
%
%
%
\begin{appendix}
%
%
\section{Irradiation at start}
\begin{figure}
\centering
\includegraphics[width=0.8\linewidth]{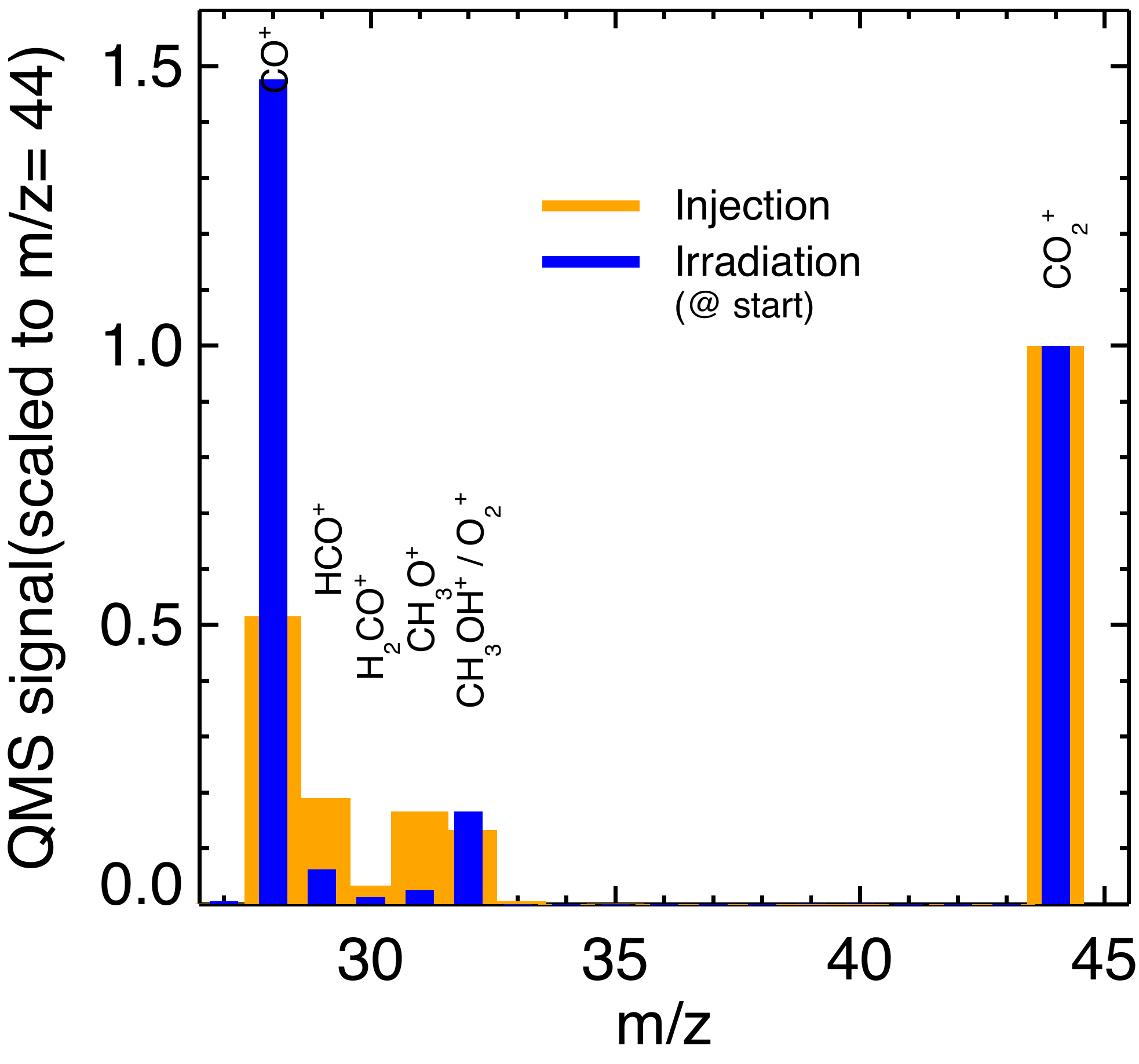}
\includegraphics[width=0.8\linewidth]{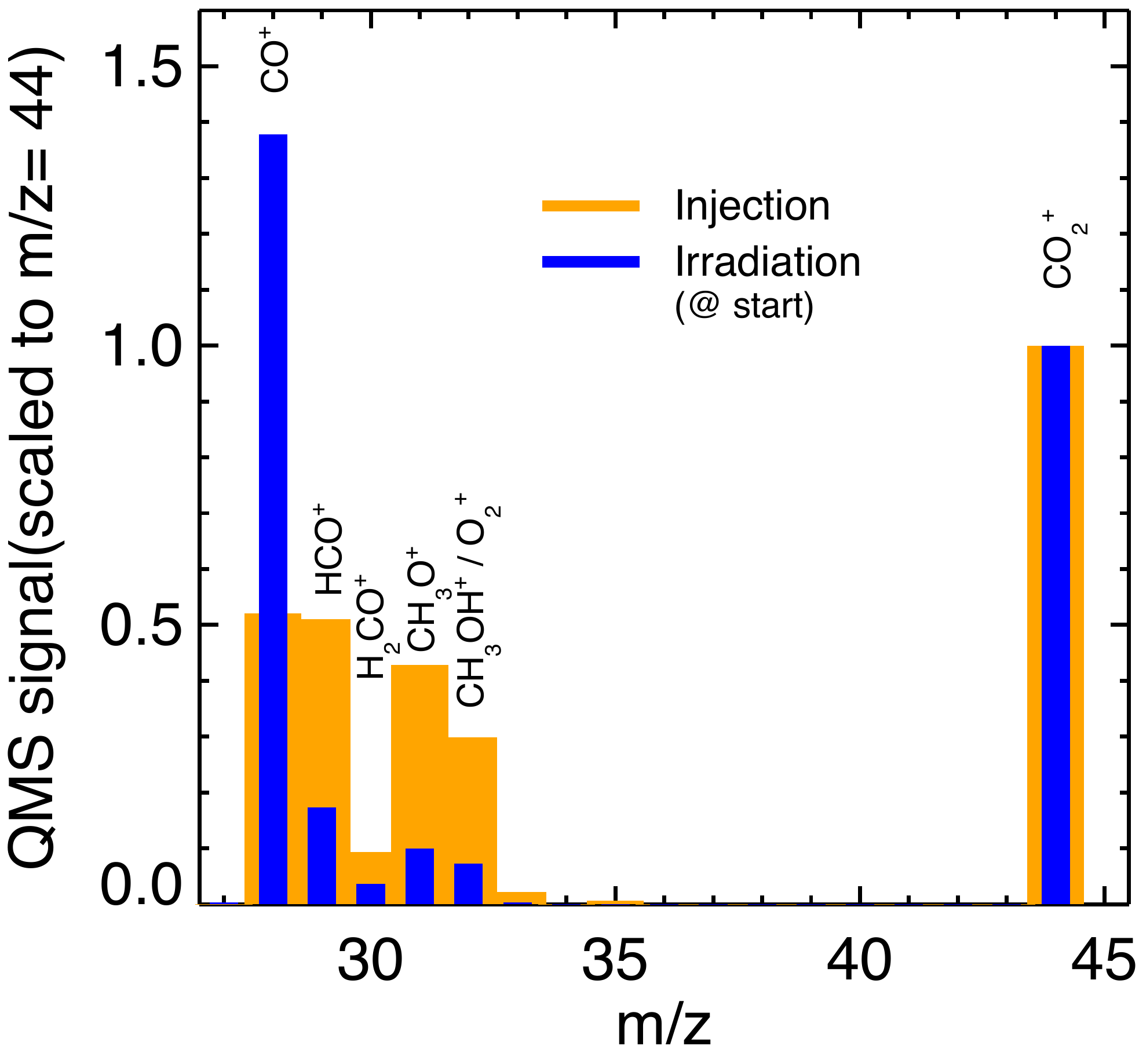}
\caption{Average of mass spectra recorded during the $\rm^{12}C-$methanol and carbon dioxide mixture injection to form the ice film (orange histogram), tracing the QMS mass-fragmentation patterns of pure methanol and carbon dioxide; average of a few spectra that were recorded at the beginning of the ice-film irradiation (blue histogram). Both mass spectra are scaled to m/z=44 (CO$_2^+$) for comparison. From top to bottom; mixtures with increasing proportions of methanol in relation to carbon dioxide. See text for details.} \label{fig:fragmentation_patterns_CH3OH_CO2}
\end{figure}

\begin{figure}
\centering
\includegraphics[width=0.8\linewidth]{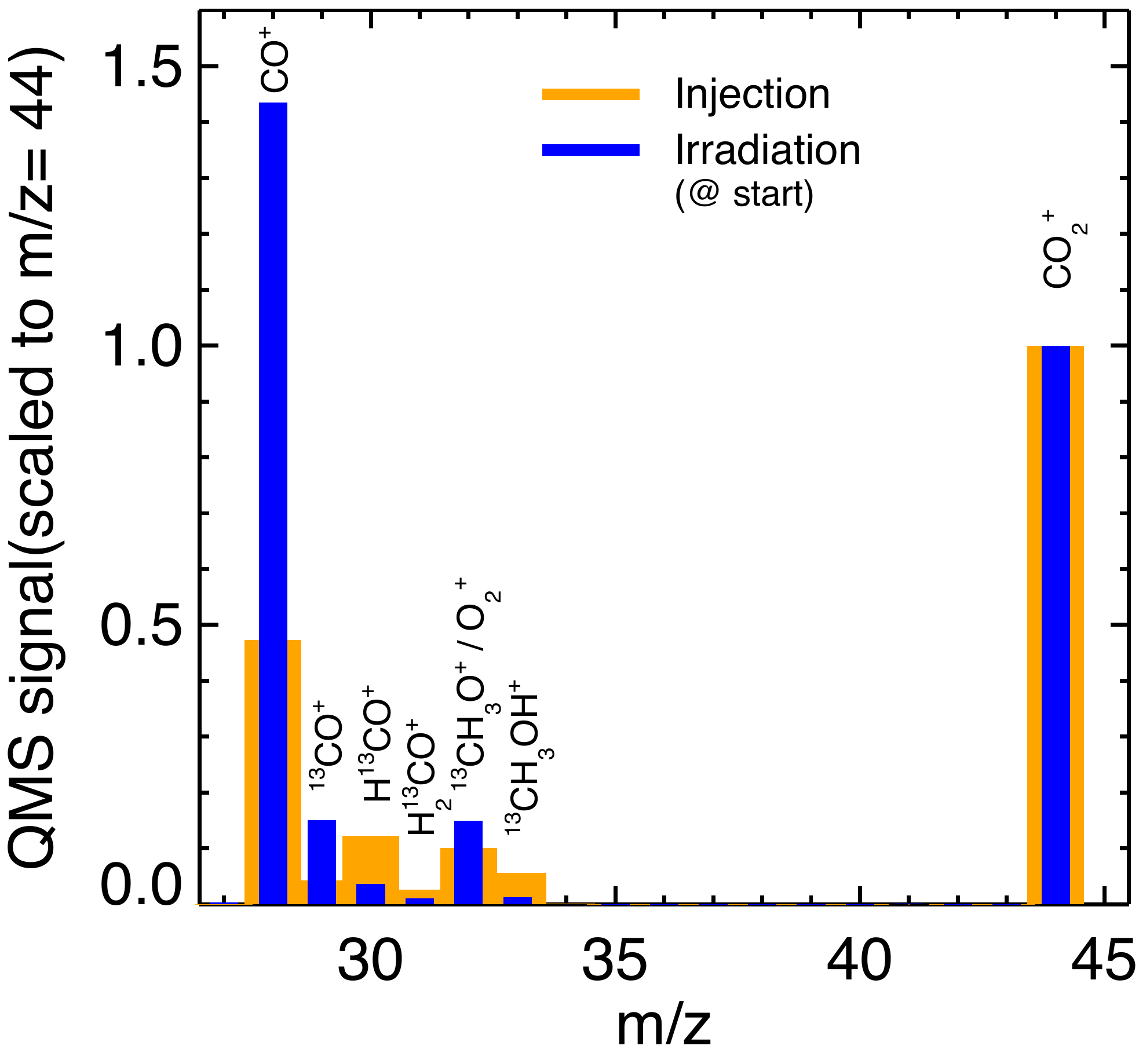}
\includegraphics[width=0.8\linewidth]{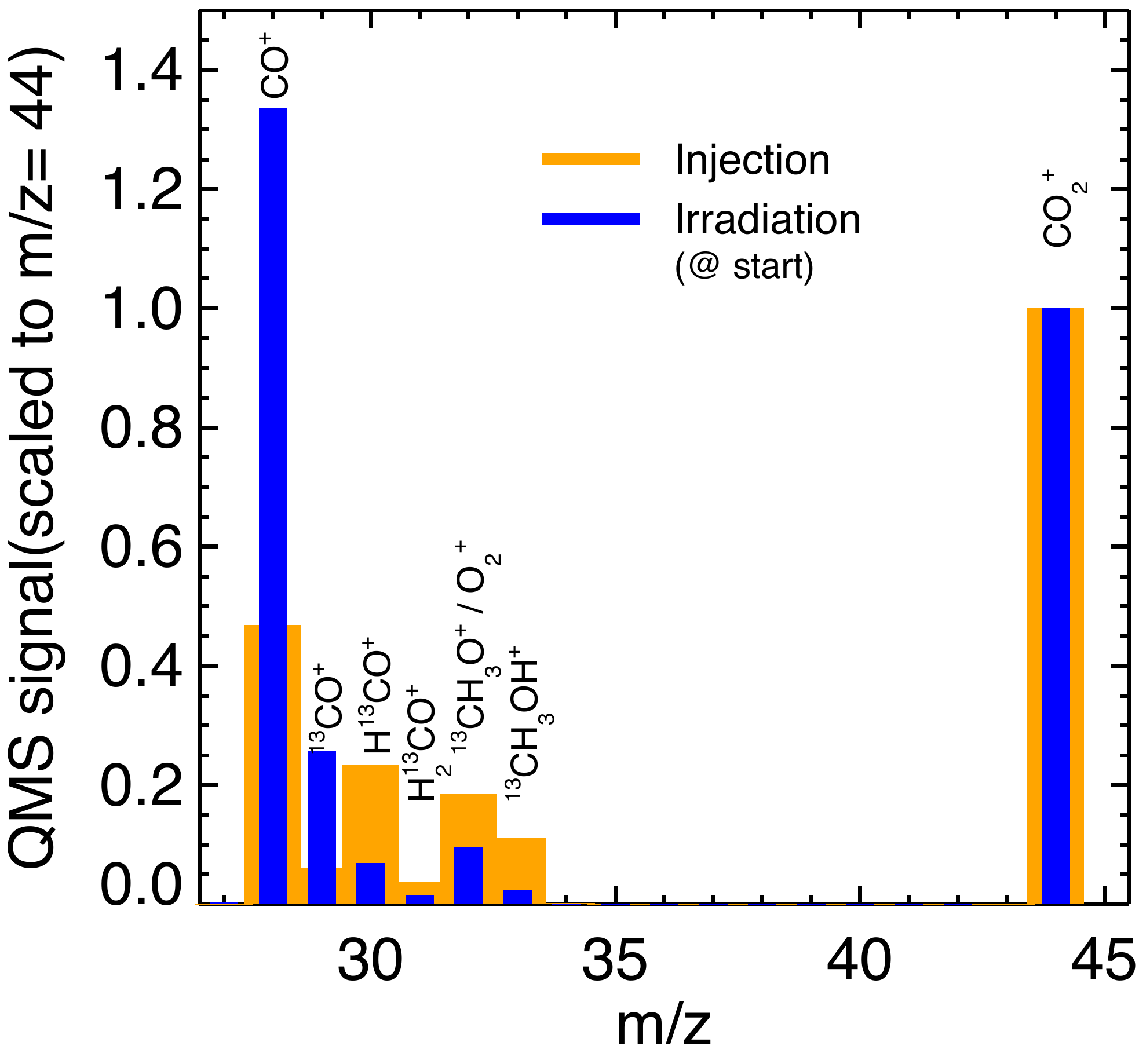}
\includegraphics[width=0.8\linewidth]{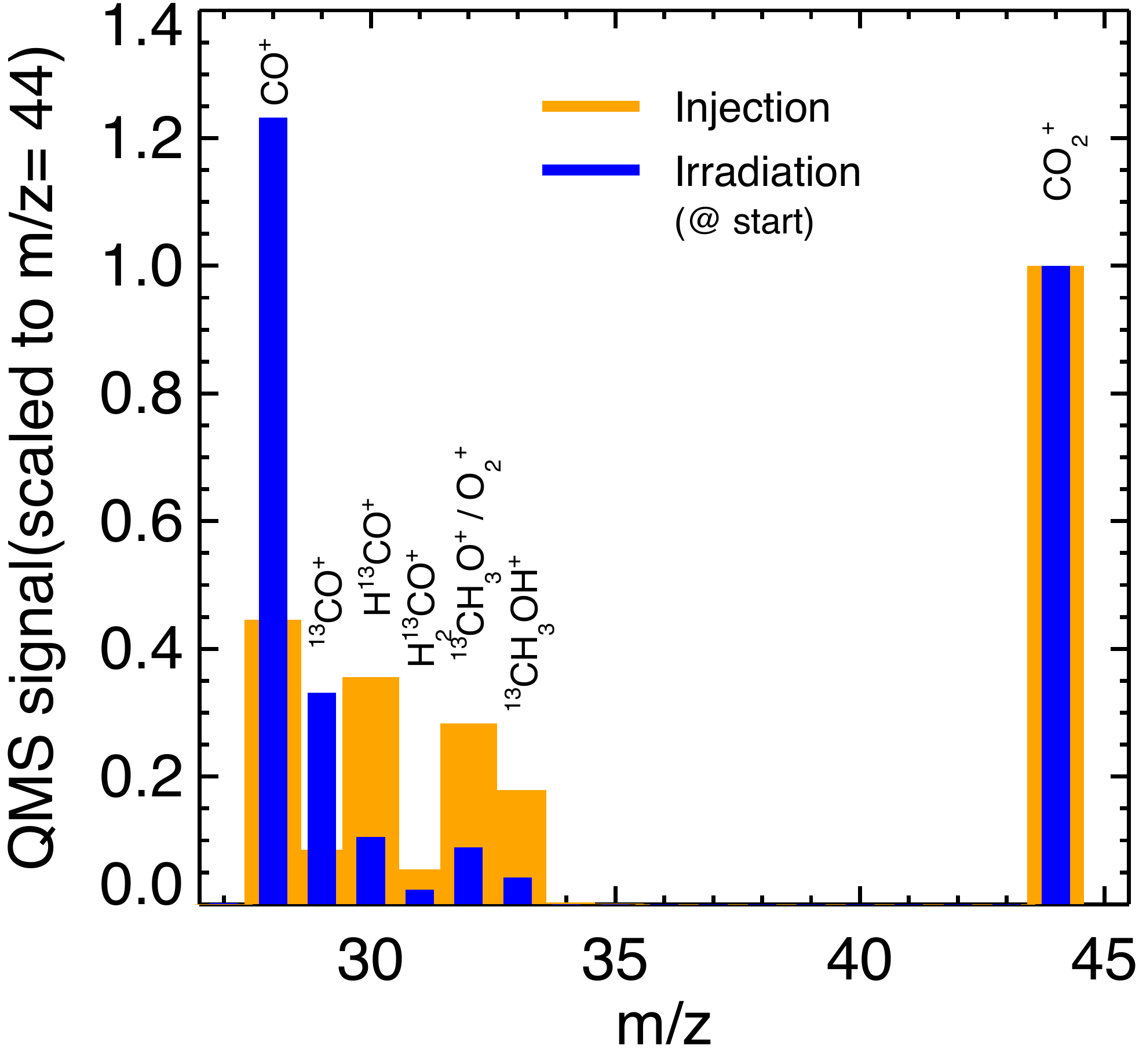}
\caption{Average of mass spectra recorded during the labelled $^{13}$C-methanol and carbon dioxide mixture injection to form the ice film (orange histogram), tracing the QMS mass-fragmentation patterns of pure methanol and carbon dioxide; average of a few spectra that were recorded at the beginning of the ice-film irradiation (blue histogram). Both mass spectra are scaled to m/z=44 (CO$_2^+$) for comparison. From top to bottom; mixtures with increasing proportions of $^{13}$C-methanol in relation to carbon dioxide. See text for details.} \label{fig:fragmentation_patterns_13CH3OH_CO2}
\end{figure}

The average of the mass spectra recorded during the methanol and carbon dioxide mixture injection are shown in Fig.\ref{fig:fragmentation_patterns_CH3OH_CO2}, as well as QMS spectra recorded at the beginning of the ion irradiation. These diagrams show that the main radiolysis product of CO$_2$ is, as expected, CO (m/z=28).  
The m/z=32 QMS channel cannot be used to follow methanol, as it is also populated by dioxygen arising from the radiolysis of carbon dioxide. Instead, m/z=31 is a good alternative before the use of labelled $^{13}$C-methanol to unambiguously separate these species in the QMS detection.
The average of the mass spectra recorded during the labelled $^{13}$C-methanol and carbon dioxide mixture injection are shown in Fig.\ref{fig:fragmentation_patterns_13CH3OH_CO2}, as well as QMS spectra recorded at the beginning of the ion irradiation. These diagrams show that
the main radiolysis product of CO$_2$ is, as expected, CO (m/z=28), and the main radiolysis product of $^{13}$C-methanol is $^{13}$CO (m/z=29).
%
%
%
\section{Infrared spectra and band strengths}
%
The infrared spectra, baseline corrected, used to evaluate the column densities of the bulk ice mixture evolution during irradiation are shown. Band strengths used in the analysis are given in Table~\ref{table:band_strengths}
\begin{figure}[tbhp]
\centering
\includegraphics[width=\linewidth]{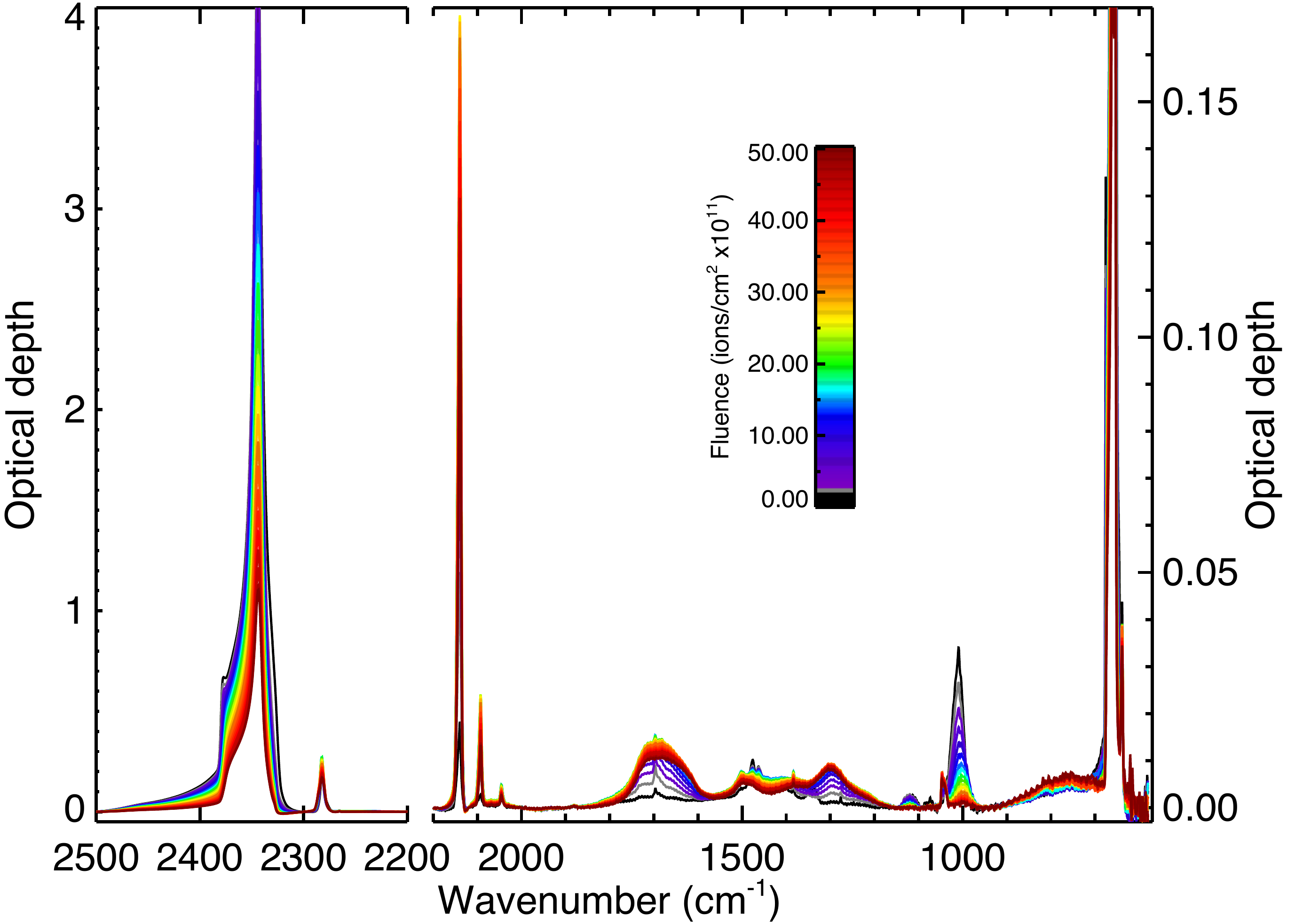}
\caption{Methanol ($\rm^{13}C-$ isotopologue) carbon dioxide ice-film mixtures experiment W1. Evolution of the infrared spectra as a function of fluence. 
The panel is split into two wavenumber ranges to maximise the optical depth scale.
}
\label{fig:IR_sputtering_190503_W1}
\end{figure}
%
%
\begin{figure}[tbhp]
\centering
\includegraphics[width=\linewidth]{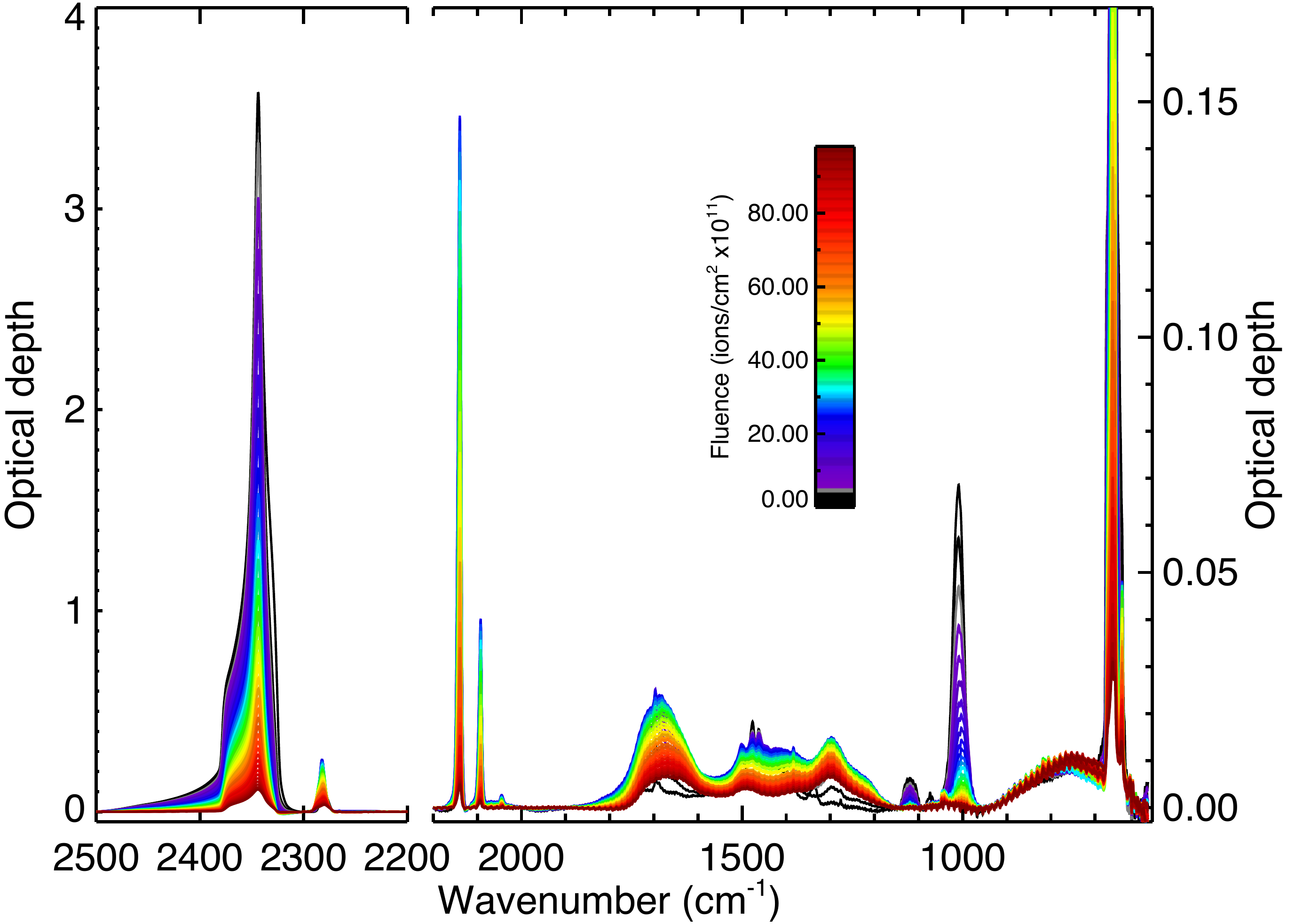}
\caption{Methanol ($\rm^{13}C-$ isotopologue) carbon dioxide ice-film mixtures experiment W1 BIS. Evolution of the infrared spectra as a function of fluence. 
The panel is split into two wavenumber ranges to maximise the optical depth scale.
}
\label{fig:IR_sputtering_190503_W1_BIS}
\end{figure}
%
%
\begin{figure}[tbhp]
\centering
\includegraphics[width=\linewidth]{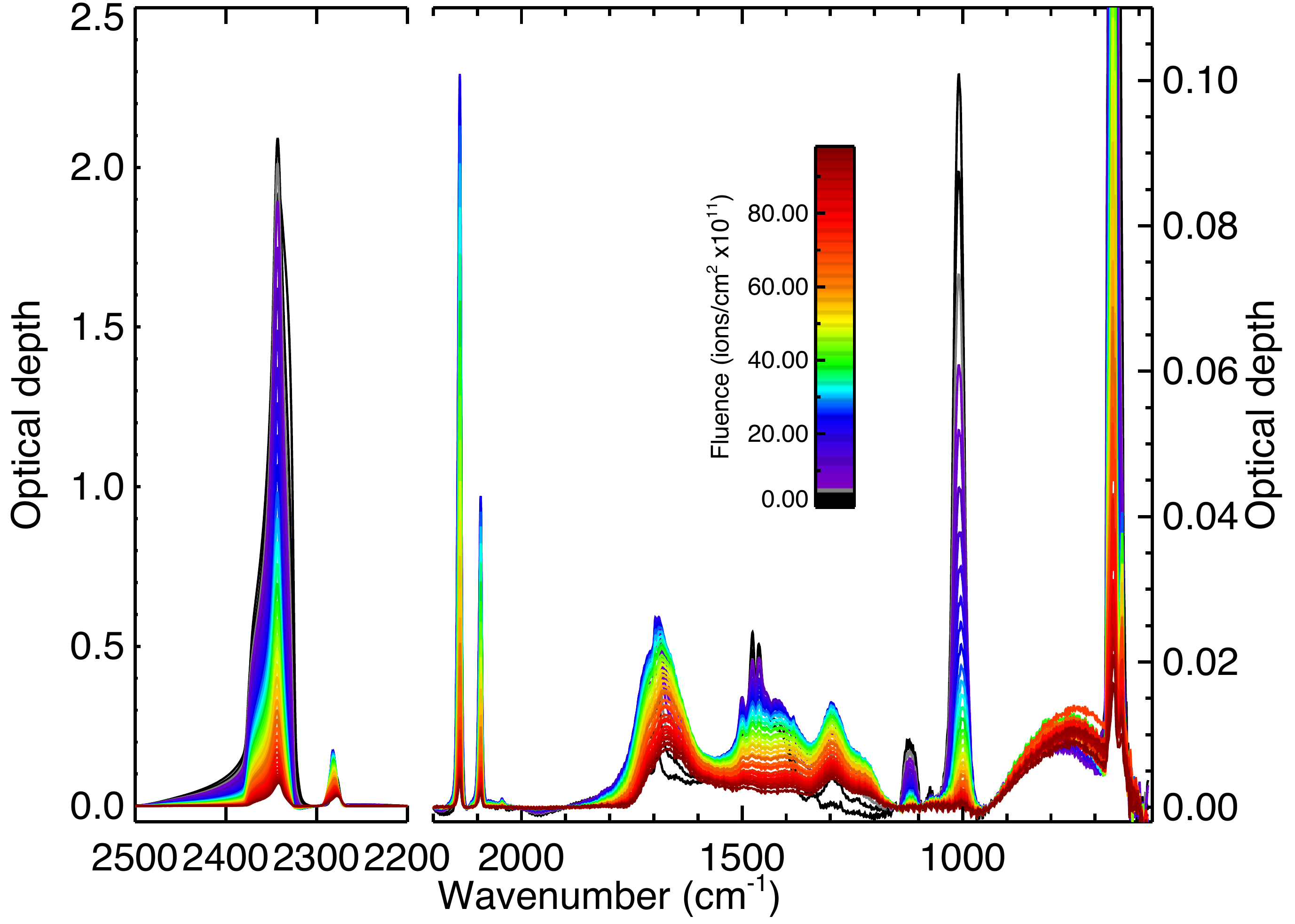}
\caption{Methanol ($\rm^{13}C-$ isotopologue) carbon dioxide ice-film mixtures experiment W1 TER. Evolution of the infrared spectra as a function of fluence. 
The panel is split into two wavenumber ranges to maximise the optical depth scale.
}
\label{fig:IR_sputtering_190503_W1_TER}
\end{figure}
%
%
\begin{figure}[tbhp]
\centering
\includegraphics[width=\linewidth]{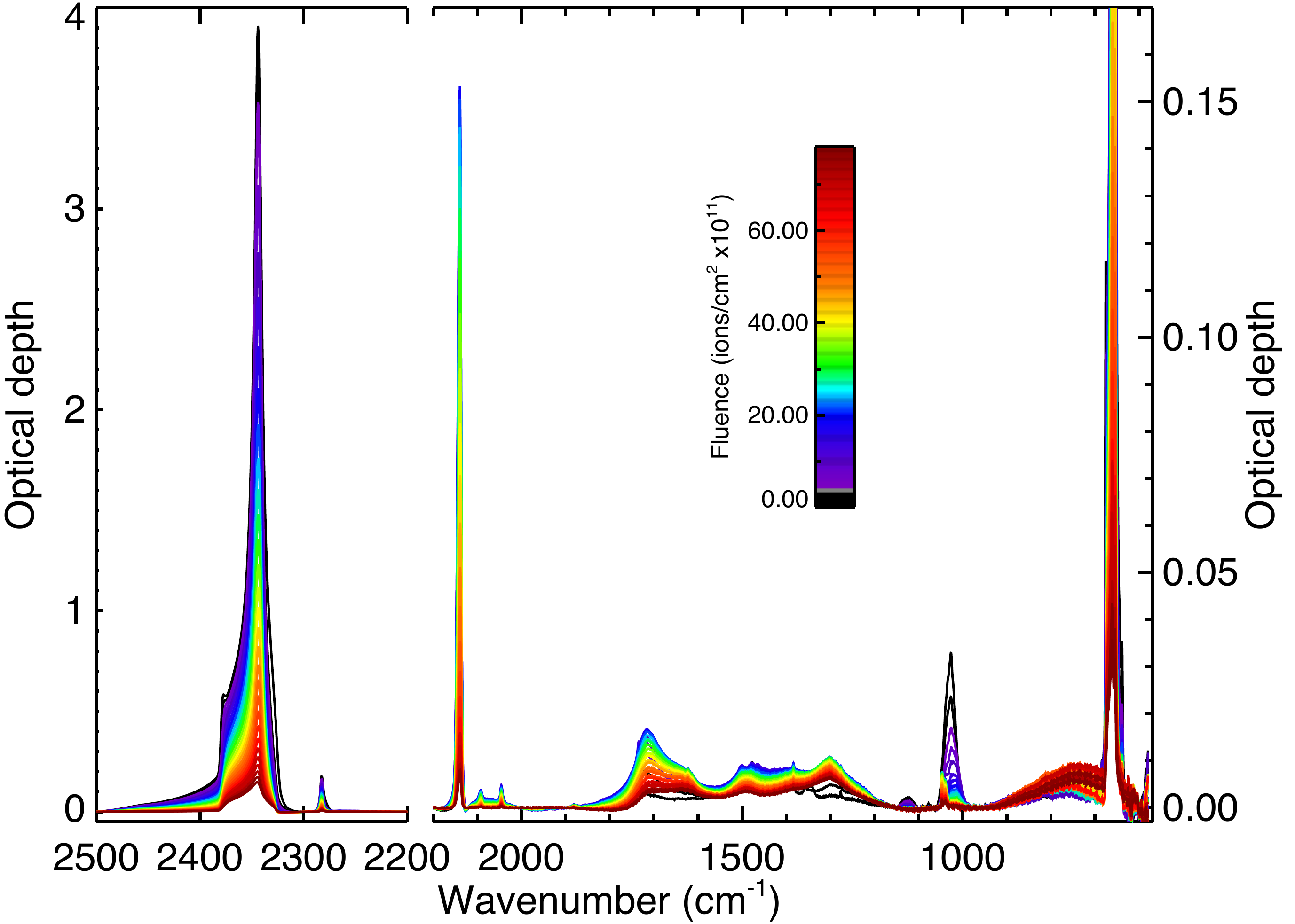}
\caption{Methanol ($\rm^{12}C-$ isotopologue) carbon dioxide ice-film mixtures experiment W2. Evolution of the infrared spectra as a function of fluence. 
The panel is split into two wavenumber ranges to maximise the optical depth scale.
}
\label{fig:IR_sputtering_190503_W2}
\end{figure}
%
%
\begin{figure}[tbhp]
\centering
\includegraphics[width=\linewidth]{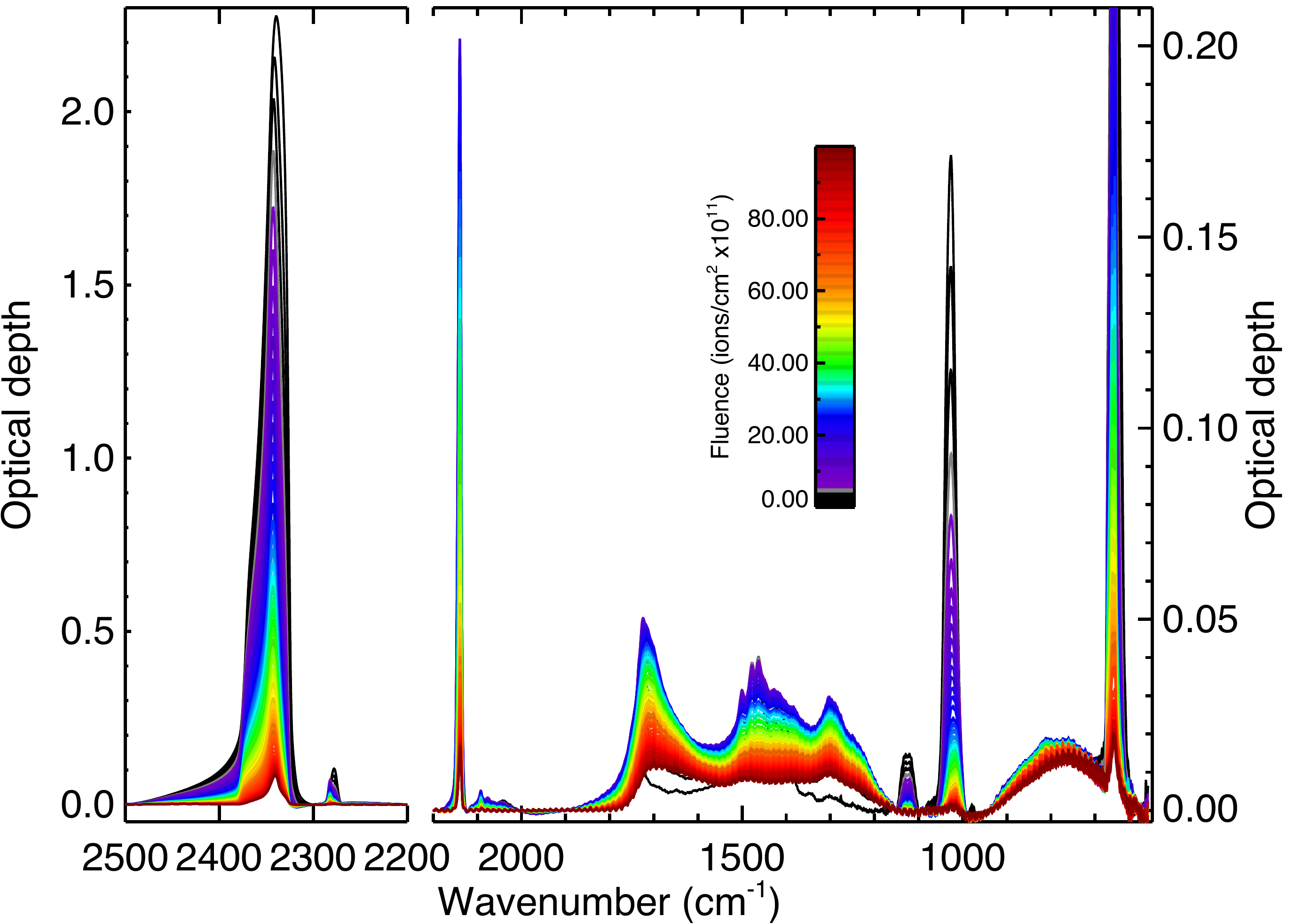}
\caption{Methanol ($\rm^{12}C-$ isotopologue) carbon dioxide ice-film mixtures experiment W3. Evolution of the infrared spectra as a function of fluence. 
The panel is split into two wavenumber ranges to maximise the optical depth scale.
}
\label{fig:IR_sputtering_190503_W3}
\end{figure}
%
%
%
\begin{table*}
\caption{Integrated band strengths used in the analysis}             
\begin{center}
\begin{tabular}{l l l l l }     
\hline\hline       
Species                         &Mode                   &Position                               &A                              &Ref            \\
                                        &                               &cm$^{-1}$                      &cm.molec$^{-1}$        &               \\
\hline
CO                                      &CO stretch             &2140                   &$\rm 1.1\times10^{-17}$                      &\cite{Jiang1975}       \\
                                        &                               &                                 &$\rm 1.1\times10^{-17}$                        &\cite{Gerakines1995}   \\
                                        &                               &                               &$\rm 1.12\times10^{-17}$             &\cite{Bouilloud2015}   \\
                                        &                               &                                 &$\rm 1.1\times10^{-17}$                        &{\bf\it Adopted for this work}          \\
CO$_2$                          &CO$_2$ stretch &2350                   &$\rm 7.6\times10^{-17}$                      &\cite{Gerakines1995} \\
CH$_3$OH                        &OH stretch             &3600-2700              &$\rm 1.1\times10^{-16}$                      &\cite{ldh1986} \\
                                        &                               &                                 &$\rm 1.28\times10^{-16}$               &\cite{Palumbo1999}             \\
                                        &                               &                                 &$\rm 1.0\times10^{-16}$                        &\cite{Bouilloud2015}           \\
                                        &                               &                                 &$\rm 1.1\pm0.15\times10^{-16}$ &{\bf\it Adopted for this work}          \\
CH$_3$OH                        &C-O stretch            &1030                   &$\rm 1.8\times10^{-17}$                      &\cite{ldh1986} \\
                                        &                               &                               &$\rm 1.8\times10^{-17}$                      &\cite{Sandford1993}    \\
                                        &                               &                               &$\rm 1.2\times10^{-17}$                      &\cite{Palumbo1999}             \\
                                        &                               &                               &$\rm 1.07\times10^{-17}$             &\cite{Bouilloud2015}           \\
                                        &                               &                                 &$\rm 1.5\pm^{0.3}_{0.4}\times10^{-17}$ &{\bf\it Adopted for this work}          \\
\hline                  
\end{tabular}
\end{center}
\label{table:band_strengths}
\end{table*}
%

%
\end{appendix}
\end{document}